\documentclass[twoside]{dis08}
\usepackage[latin1]{inputenc}
\usepackage[dvips]{graphicx,epsfig,color}
\usepackage{wrapfig,rotating}
\usepackage{amssymb,amsmath,array}

\begin{document}
\title{Jet Production in Polarized pp Collisions at RHIC}

%***********************************************************************
% AUTHORS INFORMATION AREA
%***********************************************************************
\author{C. A. Gagliardi$^1$, for the STAR Collaboration
%
% Optional short acknowledgment: remove next line if non-needed
%
% DO NOT MODIFY THE FOLLOWING '\vspace' ARGUMENT
\vspace{.3cm}\\
%
% Addresses and institutions (remove "1- " in case of a single institution)
1- Cyclotron Institute, Texas A\&M University \\
College Station, Texas, 77843  U.S.A.
%
% Remove the next three lines in case of a single institution
}
%***********************************************************************
% END OF AUTHORS INFORMATION AREA
%***********************************************************************

\maketitle

\begin{abstract}
The STAR Collaboration has measured the longitudinal double-spin
asymmetry for inclusive jet production in polarized p+p collisions
at $\sqrt{s}$ = 200 GeV.  The results set significant new constraints
on the gluon polarization within the nucleon.  Future measurements
of asymmetries for di-jet production will provide direct access to
the momentum dependence of the gluon polarization, $\Delta g(x,Q^2)$.
\end{abstract}

\section{Introduction}

One of the primary goals of the RHIC spin program is to determine the gluon polarization distribution within the proton.  At leading order, $pp$ collisions at RHIC involve a mixture of quark-quark ($qq$), quark-gluon ($qg$), and gluon-gluon ($gg$) scattering processes, with $qg$ and $gg$ contributions dominating.  This makes RHIC an ideal place to study gluon polarization.  For the past several years, the STAR Collaboration has been studying mid-rapidity inclusive jet production in $\sqrt{s}$ = 200 GeV $pp$ collisions in order to constrain the gluon contribution to the proton spin.  We find that the cross section for inclusive jet production is well described by next-to-leading-order (NLO) perturbative QCD predictions for jet transverse momenta ($p_T$) spanning the range 5 to 50 GeV/c \cite{Runs34jets}.  Our first measurements of the longitudinal double-spin asymmetry, $A_{LL}$, for inclusive jet production \cite{Runs34jets}, based on data that were recorded during the 2003 and 2004 RHIC runs, disfavored a maximal gluon polarization scenario where the gluon polarization distribution, $\Delta g(x,Q_0^2)$, at the input scale, $Q_0^2$ = 0.4 GeV$^2$, is assumed to be equal to the unpolarized gluon distribution, $g(x,Q_0^2)$.

Here, we describe measurements of $A_{LL}$ for inclusive jet production that were performed by STAR during the 2005 \cite{Run5jets} and 2006 RHIC runs.  The increased precision and kinematic coverage of these data provide valuable new constraints on the gluon polarization in the proton when included in a NLO global analysis.  We also describe the STAR plans to study di-jet production in upcoming RHIC runs.  This will provide direct leading-order access to the momentum dependence of the gluon polarization, $\Delta g(x,Q^2)$.

\section{Inclusive jet asymmetries}

\begin{figure}[t]
  \begin{minipage}{0.49\columnwidth}
    \includegraphics*[width=\textwidth]{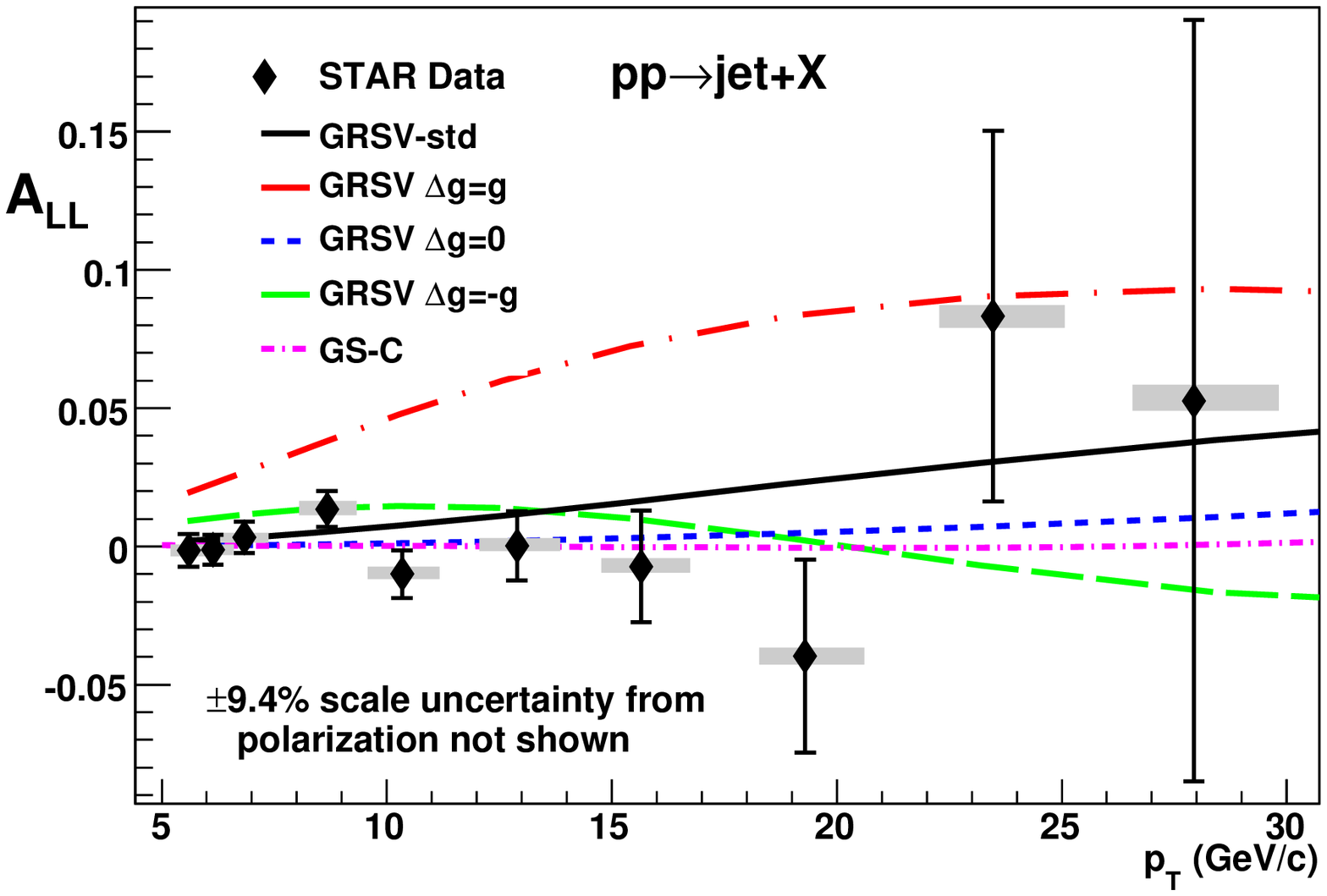}
  \end{minipage}
  \hfill
  \begin{minipage}{0.49\columnwidth}
    \includegraphics*[width=\textwidth]{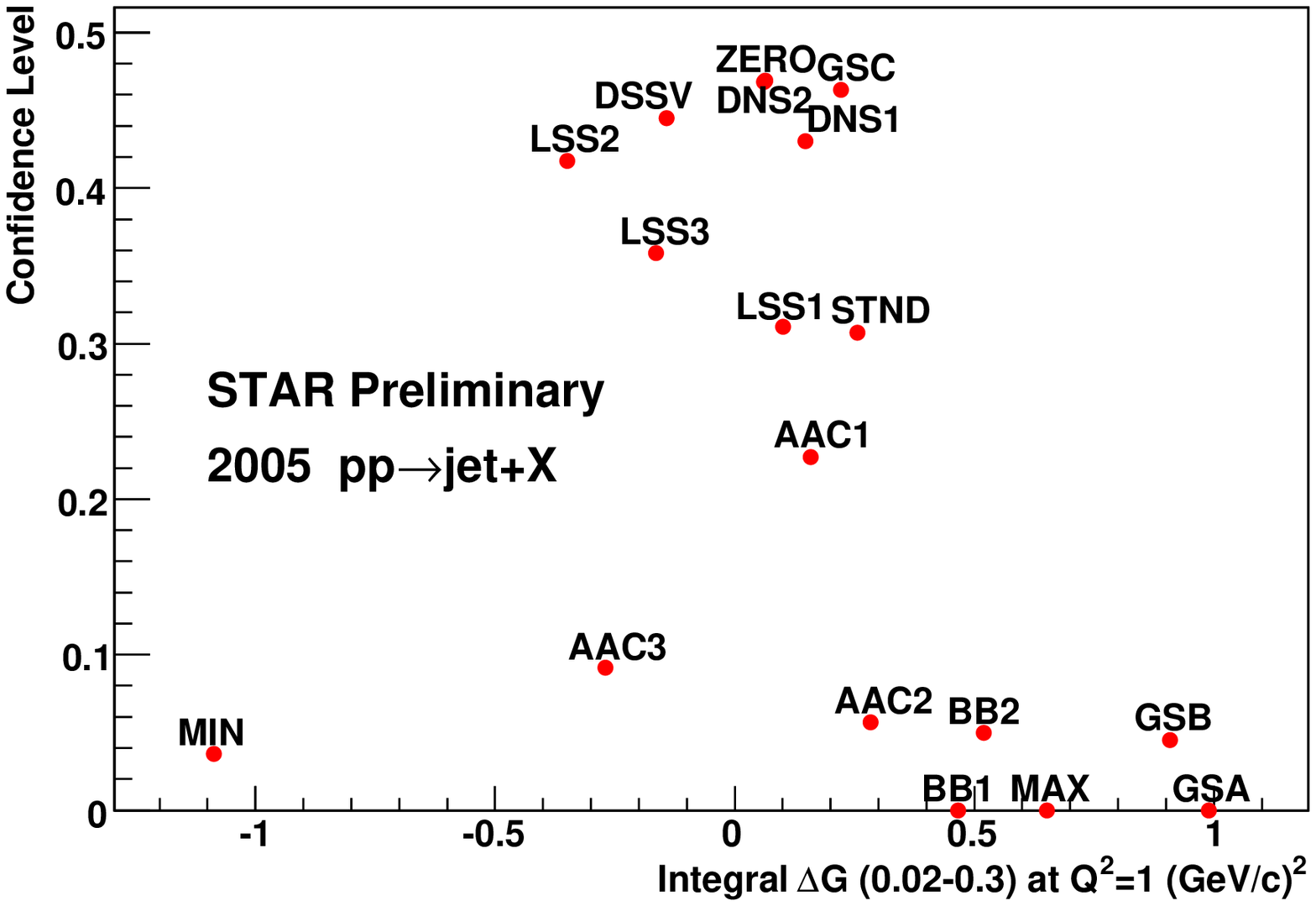}
  \end{minipage}
\caption{Left:  $A_{LL}$ vs.\@ $p_T$ for inclusive jet production at $0.2<\eta<0.8$ and $\sqrt{s}$ = 200 GeV, based on data recorded during 2005 \protect\cite{Run5jets}.  The error bars are statistical; the gray bands show the experimental systematic uncertainties.
Right:  Confidence level calculations that various global analyses of polarized DIS measurements are consistent with the $A_{LL}$ results.}\label{Fig:05}
\end{figure}

STAR measures jets with the time projection chamber (TPC) plus the barrel and endcap electromagnetic calorimeters (BEMC/EEMC).  The TPC provides tracking information for charged particles; the calorimeters provide triggering and detection of photons and electrons.  Jets are reconstructed using a mid-point cone algorithm, as described in \cite{Runs34jets}.  In 2005, the BEMC covered a pseudorapidity range $0<\eta<1$, and EEMC information was not considered in the jet reconstruction.  To minimize the effects of the BEMC acceptance on the jet energy scale, the cone radius was limited to $R = \sqrt{\Delta\eta^2 + \Delta\phi^2} = 0.4$, and the jet axis was required to point between $0.2<\eta<0.8$.  In 2006, the BEMC covered $-1<\eta<1$, and the EEMC information was included in the jet reconstruction.  These changes permitted the cone radius to be increased to $R=0.7$, thereby increasing the jet reconstruction efficiency and decreasing the reconstruction biases.  The allowed range for the jet axis was expanded to $-0.7<\eta<0.9$.  Events were recorded if they satisfied either a high tower (a minimum energy deposition in a $\Delta\eta \times \Delta\phi = 0.05 \times 0.05$ tower) or jet patch (a minimum energy deposition over a $\Delta\eta \times \Delta\phi = 1 \times 1$ region) trigger.  Most of the events for 2005, and all of the events presented here for 2006, were derived from the jet patch trigger.

The polarizations of the beams were determined from measurements of $p$+C Coulomb-nuclear interference \cite{CNI} calibrated via a polarized atomic hydrogen gas-jet target \cite{gasjet}.  The helicities of the two colliding beams varied from bunch to bunch; the helicity pattern was changed from fill to fill.  Relative luminosities for the various spin states were measured with beam-beam counters located up and downstream of the STAR interaction region.

The dominant systematic uncertainties in the $A_{LL}$ measurements arise from trigger and reconstruction biases.  These lead to shifts between the true and observed jet energies, which have been corrected in the results presented here.  They can also introduce distortions in the measured $A_{LL}$ values that depend on the underlying form of the gluon polarization distribution, which have been estimated as described in \cite{Run5jets}.

\begin{figure}[t]
  \begin{minipage}{0.49\columnwidth}
    \includegraphics*[width=\textwidth]{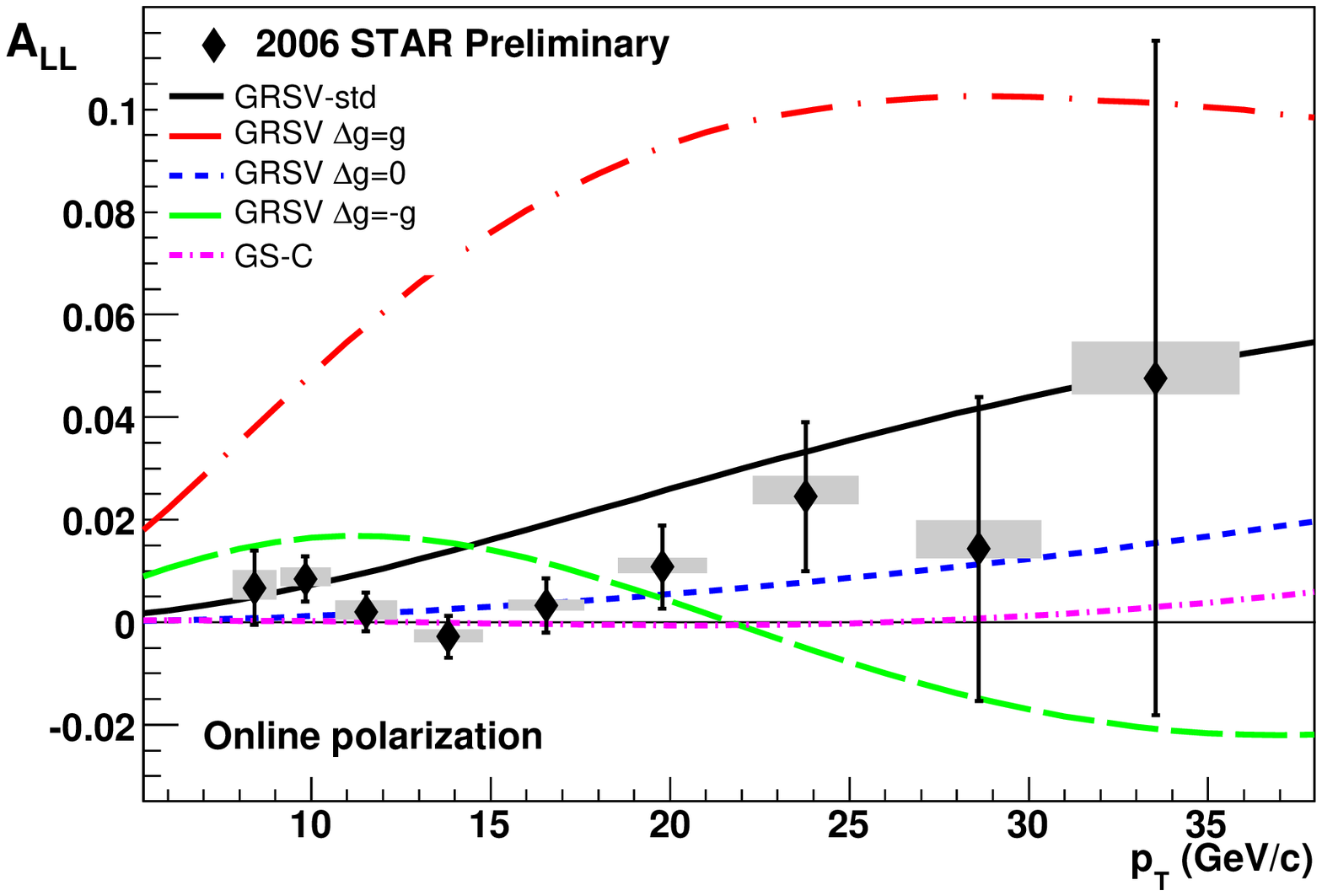}
  \end{minipage}
  \hfill
  \begin{minipage}{0.49\columnwidth}
    \includegraphics*[width=\textwidth]{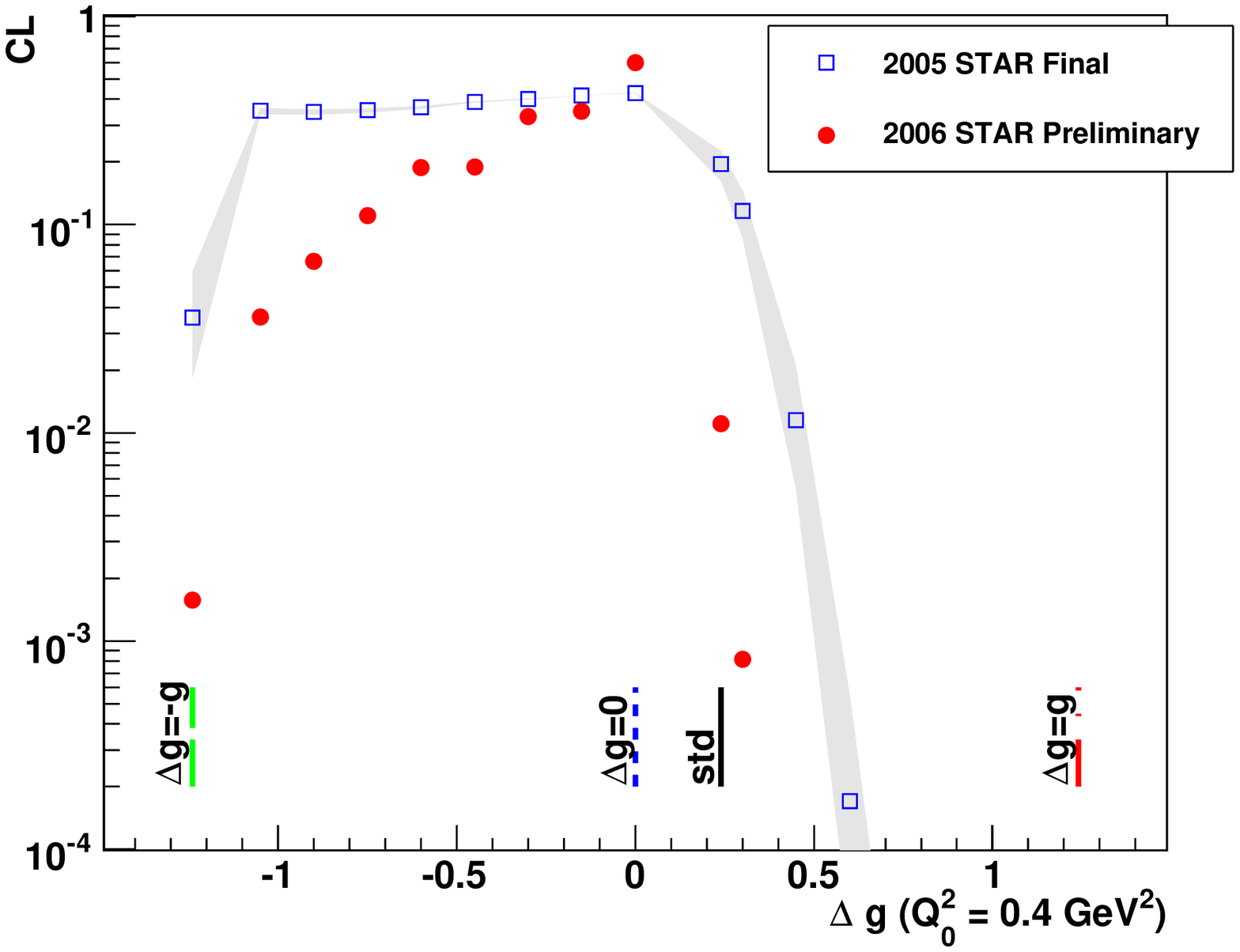}
  \end{minipage}
\caption{Left:  Preliminary STAR $A_{LL}$ vs.\@ $p_T$ for inclusive jet production at $-0.7<\eta<0.9$ and $\sqrt{s}$ = 200 GeV, based on data recorded during 2006.  The error bars are statistical; the gray bands show the experimental systematic uncertainties.  Right:  Confidence level calculations that various global fits in the GRSV framework having different gluon polarizations are consistent with the $A_{LL}$ results from 2005 and 2006.}\label{Fig:06}
\end{figure}

Figure \ref{Fig:05} shows the $A_{LL}$ results for 2005 \cite{Run5jets}, in comparison to predictions from five fits to previous polarized inclusive deep-inelastic scattering (DIS) measurements.  GRSV-std represents an overall best fit to the DIS data, while GRSV $\Delta g=g$, $\Delta g=0$, and $\Delta g=-g$ illustrate fits with constrained gluon polarization distributions at the input scale \cite{GRSV}.  These fits are representative of most recent global analyses.  In contrast, GS-C has a large positive gluon polarization at low $x$, a node near $x \sim 0.1$, and a negative gluon polarization at large $x$ \cite{GSC}.  The GRSV $\Delta g=g$ fit is not consistent with the data.  Many other recent global analyses also predict large values of $A_{LL}$ \cite{url} and are excluded by the present results, as shown in the right panel of Fig.\@ \ref{Fig:05}.

Figure \ref{Fig:06} shows preliminary $A_{LL}$ results for 2006, in comparison to predictions from the same five global analyses.  The increased acceptance of the STAR detector in 2006, combined with increases in the integrated luminosity and beam polarization, lead to statistical uncertainties that are a factor of 3-4 times smaller for $p_T>13$ GeV/c than in the 2005 data.  The higher precision provides even tighter constraints on the gluon polarization in the proton.  This is illustrated by the right panel of Fig.\@ \ref{Fig:06}, where confidence levels have been calculated based on a series of global fits within the GRSV framework with fixed values of the integral gluon polarization at the input scale \cite{MarcoWerner}.  The new measurements particularly constrain negative values of gluon polarization.  For example, there is $<$\,0.2\% probability that the GRSV $\Delta g=-g$ scenario is consistent with the 2006 results.

Very recently, these inclusive jet asymmetries have been included in the DSSV global analysis \cite{DSSV}.  This is the first NLO global analysis to treat inclusive DIS, semi-inclusive DIS, and $pp$ collision data on an equal basis.  This new analysis includes a node near $x \sim 0.1$ in the gluon polarization distribution for $Q^2 < 10$ GeV$^2$, which evolves away at higher scales, but with the opposite phase compared to GS-C.  A detailed study of the $\chi^2$ distribution \cite{DSSV} shows that these STAR data provide the strongest limits on negative gluon polarization over the range $0.05<x<0.2$ and also contribute significantly to the limits on positive gluon polarization over the same range.

\begin{figure}[t]
  \begin{minipage}{0.72\columnwidth}
    \includegraphics*[bb=0 0 567 548,width=\textwidth]{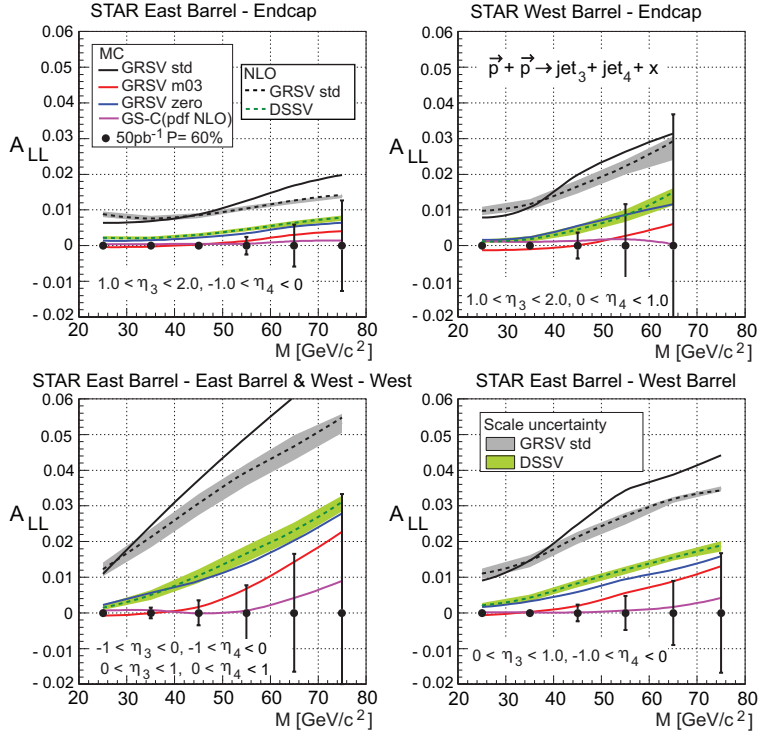}
  \end{minipage}
  \hfill
  \begin{minipage}{0.26\columnwidth}
    \caption{Sensitivity of STAR di-jet $A_{LL}$ measurements for the upcoming RHIC run, compared to PYTHIA-based calculations of the expected asymmetries for 4 gluon polarization models and full NLO calculations of the expected asymmetries, including uncertainties due to scale variations (bands), for GRSV-std and DSSV.  The projected uncertainties assume an integrated p+p luminosity of 50 pb$^{-1}$ at $\sqrt{s}$ = 200 GeV with 60\% beam polarization.}\label{Fig:09}
  \end{minipage}
\end{figure}

\section{Future measurements}

Inclusive jet asymmetries at any given $p_T$ provide sensitivity to gluon polarization over a broad range of $x$ values, as illustrated in \cite{Run5jets}.  This can hide considerable structure if $\Delta g(x,Q^2)$ has a node, as in GS-C or DSSV.  STAR is embarking on measurements of $A_{LL}$ for di-jet production to obtain more detailed information about the gluon polarization distribution.  At leading order, the momentum fractions of the incident partons can be calculated from the mass, $M$, and rapidity, $y$, of the di-jet via $x_{1,2} = (M/\sqrt{s})e^{\pm y}$.  The large acceptance of the STAR detector provides the additional benefit that the di-jet kinematics can be chosen to emphasize the scattering of low-$x$ gluons by highly polarized valence quarks.

Figure \ref{Fig:09} shows the expected precision of the STAR di-jet measurements during the upcoming RHIC run.  The four panels provide a rough separation of the events according to di-jet rapidity and partonic scattering angle.  The curves show PYTHIA-based predictions for the asymmetries based on GRSV-std, GRSV $\Delta g = 0$, GRSV m03 (a fit in the GRSV framework with an integral of the gluon polarization equal to $-0.3$ at the input scale), and GS-C.  The bands show full NLO calculations of the predicted asymmetries for GRSV-std and DSSV.  The power of these measurements can be illustrated by considering the first three data points of the lower-left panel.  They will determine $\Delta g(x)/g(x)$ to approximately $\pm$0.02-0.03 in three $x$ bins centered at 0.076, 0.10, and 0.13.  The first three points in the upper-left panel will also sample $\Delta g(x)/g(x)$ at the same three $x$ values, but with somewhat reduced precision.  Overall, this measurement will determine $\Delta g(x,Q^2)$ over the range $0.04<x<0.3$.  In the further future, STAR will study inclusive jet and di-jet production in $pp$ collisions at $\sqrt{s}$ = 500 GeV, which will expand the kinematic reach to even lower-$x$ gluons.

\begin{footnotesize}
% IF YOU DO NOT USE BIBTEX, USE THE FOLLOWING SAMPLE SCHEME FOR THE REFERENCES
% ----------------------------------------------------------------------------

% ----------------------------------------------------------------------------

% IF YOU USE BIBTEX,
% - DELETE THE TEXT BETWEEN THE TWO ABOVE DASHED LINES
% - UNCOMMENT THE NEXT TWO LINES AND REPLACE 'Name_Of_Your_BibFile'

%\bibliographystyle{unsrt}
%\bibliography{Name_Of_Your_BibFile}
% example of Name_Of_Your_BibFile.bib
% @Article{Turcato:2006ch,
%      author    = "Turcato, M.",
%  collaboration = "ZEUS and H1",
%      title     = "Lepton flavour violation and charmonium physics at HERA",
%      journal   = "Nucl. Phys. Proc. Suppl.",
%      volume    = "162",
%      year      = "2006", 
%      pages     = "283-287",
%      SLACcitation  = "%%CITATION = NUPHZ,162,283;%%"
% }
% 
% @Unpublished{Gogitidze:2007du,
%      author    = "Gogitidze, N.",
%  collaboration = "H1", 
%      title     = "Prompt photons and particle momentum distributions at
%                   HERA", 
%      year      = "2007",
%      note    = "hep-ex/0701033",
%      SLACcitation  = "%%CITATION = HEP-EX 0701033;%%"
% }

\end{footnotesize}

% ****************************************************************************
% END OF BIBLIOGRAPHY AREA
% ****************************************************************************

\end{document}